\documentclass[floats,aps,prl,showpacs,twocolumn,superscriptaddress]{revtex4}
\usepackage{graphicx}
\usepackage{amsmath,amssymb,latexsym,amsfonts,subfigure,bbm}
\def\taub{{\boldsymbol\tau}}
\def\Sb{{\boldsymbol S}}
\begin{document}
\title{Interacting Quantum Dot Coupled to a Kondo Spin: A Universal
   Hamiltonian Study}

\author{Stefan Rotter}
\affiliation{Department of Applied Physics, Yale University, New Haven,
CT 06520, USA}
\author{Hakan E.~T\"ureci}
\affiliation{Institute of Quantum Electronics, ETH-Z\"urich, CH-8093 Z\"urich,
Switzerland}
\author{Y.~Alhassid}
\affiliation{Center for Theoretical Physics, Sloane Physics Laboratory, Yale
     University, New Haven, CT 06520, USA}
\author{A.~Douglas Stone}
\affiliation{Department of Applied Physics, Yale University, New Haven,
CT 06520, USA}
\affiliation{Center for Theoretical Physics, Sloane Physics Laboratory, Yale
     University, New Haven, CT 06520, USA}

\date{\today}

\begin{abstract}
We study a mesoscopic interacting quantum dot described by the
``universal Hamiltonian" that is coupled to a
Kondo spin. The ferromagnetic exchange interaction within the dot leads to a
stepwise increase of the ground state spin; this Stoner staircase is
modified non-trivially by the Kondo interaction.  We find that the
spin-transition steps move to lower values of the exchange coupling
for weak Kondo interaction, but shift back up for sufficiently strong
Kondo coupling. The problem is solved numerically by diagonalizing
the system Hamiltonian in a customized good-spin basis and
analytically in the weak and strong Kondo coupling limits. The interplay
of Kondo and ferromagnetic exchange can be probed with experimentally
tunable parameters.
\end{abstract}


\maketitle

A singly-occupied localized electron level (an ``impurity''
spin) interacting with a delocalized electron gas is a paradigmatic 
system in quantum
many-body physics. It gives rise to the non-perturbative
\emph{Kondo effect} in which the localized electron's magnetic moment is fully
screened by the delocalized electrons below the Kondo temperature
$T_K$. The Kondo effect was well-studied in the context of an impurity moment
embedded in a bulk metal \cite{hewson93}. Recently, Kondo physics has been the
subject of much renewed interest \cite{leos}, following its observation in
experimentally tunable quantum dots \cite{goldhaber98}.

\begin{figure}[!b]
\includegraphics[angle=0,width=85mm]{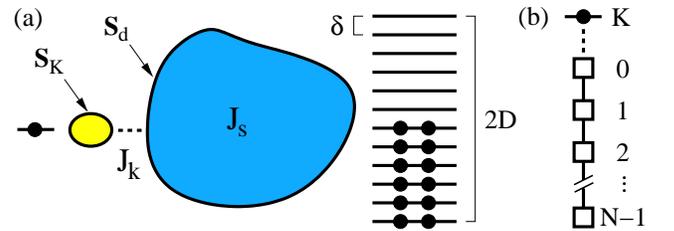}
\caption{(Color online) (a) Schematic diagram of a small quantum dot with spin
${\bf S}_K$ (Kondo spin) that is coupled antiferromagnetically (coupling
constant $J_k$) to a large quantum dot with spin ${\bf S}_d$. The large
dot, described by the universal Hamiltonian, is characterized by a
ferromagnetic exchange interaction (coupling constant $J_s$). We
assume the large dot to have $N$ equally-spaced (spacing $\delta$)
single-particle levels in a band of width $2D$ (half-filling). (b)
The large dot is represented in the site basis (squares), in which
${\bf S}_K$ couples only to site 0.}
\label{fig:1}
\end{figure}

These experimental advances have been accompanied by progress in the
theoretical treatment of the mesoscopic Kondo problem
\cite{glazman88,thimm99,simon02,kaul05,murthy05,kaul06,sindel07}. In these
mesoscopic systems, either the electrons in the leads or the electrons
in a large dot play the role of the ``electron gas" and a small
spin-1/2 dot represents the Kondo spin. Here we focus on the latter
case [see Fig.~\ref{fig:1}(a)], where the discrete single-particle
level spacing $\delta$ and mesoscopic fluctuations of the large dot
may alter the Kondo effect when $T_K \sim \delta$
\cite{thimm99,simon02,kaul05,murthy05,kaul06}.

A formidable challenge for both mesoscopic and bulk Kondo theory is to take
into account electron-electron interactions in the electron gas.
In the mesoscopic case this task is simplified when the
electron gas is confined to a large quantum dot in which the electron dynamics
is chaotic \cite{jalabert,al00}.
When the dot's Thouless energy $E_T$ is large compared with $\delta$,
the effects of electron-electron interaction are captured by the
so-called {\it universal Hamiltonian} (UH) \cite{uniham}, valid in an
interval $E_T$ around the dot's Fermi energy $E_F$. For a fixed
number of electrons, the dominant interaction in the UH is a
ferromagnetic exchange interaction.  Detailed comparison between
theory and
experiment for the statistics of Coulomb blockade peak heights and spacings
shows that including the UH ferromagnetic exchange term is both necessary and
sufficient to obtain quantitative agreement \cite{rupp}.
Thus, one can use this UH to obtain an experimentally-relevant
description of a large interacting
dot (henceforth called the ``dot'') that is Kondo-coupled
to a small dot with odd electron occupancy (henceforth called the
``Kondo spin'' ${\bf S}_K$). Such a model was first discussed in the
framework of a  mean-field approximation \cite{murthy05}, where Kondo
correlations in a dot close to its ferromagnetic Stoner instability,
$J_s\sim\delta$, were investigated. A regime just below the
instability was identified
where the Kondo coupling substantially reduces the dot's polarization. In
contrast, studies in the bulk \cite{larkin72} found
that a Kondo impurity enhances the polarization of a surrounding gas 
of electrons
at  similar high values of $J_s\lesssim\delta$.

The simplicity of the UH allows one to look for signatures of the Kondo
interaction in the magnetic properties of the dot without necessarily assuming
that the dot is very close to its bulk Stoner transition (in a
quantum dot, $J_s$ is typically a fraction of $\delta$
\cite{uniham,rupp}). Standard numerical methods for the Kondo problem
such as quantum Monte Carlo \cite{kaul06} and numerical
renormalization group (NRG) techniques \cite{sindel07} are, however,
not easily applied to this problem because the ferromagnetic exchange
coupling introduces a sign problem in the former and non-local
correlations in the latter. Here we use a customized diagonalization
method that takes advantage of the global spin rotation invariance,
using the good-spin eigenstates of the UH as a basis
\cite{rupp,tureci06}.

The Hamiltonian of our system, schematically illustrated in
Fig.~\ref{fig:1}, is given by~\cite{uniham,murthy05},
\begin{equation}
H=\sum_{\mu,\sigma=\pm} \varepsilon^{{\phantom \dagger}}_\mu\,
c^\dagger_{\mu\sigma}c^{{\phantom \dagger}}_{\mu\sigma}\!
-\!J_s\,{\bf S}_d^2\!+\!J_k\,{\bf S}_K\cdot{\bf s}_d(0)\,.
\label{eq:1}
\end{equation}
The first two terms in (\ref{eq:1}) constitute the UH of the dot \cite{uniham}
(ignoring a constant charging energy and a
Cooper channel term), described by $N$ spin-degenerate single-particle
levels $\varepsilon_\mu$ and spin $\Sb_d =
\frac{1}{2}\sum_{\mu \sigma\sigma'} c^\dagger_{\mu\sigma}
\taub_{\sigma\sigma'}c_{\mu\sigma'}$ ($\taub$ are Pauli matrices). The
coupling of the dot to the Kondo spin $\Sb_K$
(with $S_K=1/2$) is mediated by its spin density
${\bf s}_d(0)=\frac{1}{2}\sum_{\mu\nu\sigma\sigma'}\phi_\mu(0)
c^\dagger_{\mu\sigma} \taub_{\sigma\sigma'}\phi^*_{\nu}(0)c_{\nu\sigma'}$ at
the tunneling position ${\bf r}=0$ [$\phi_\mu({\bf r})$ is the orbital wave
     function of level $\mu$]. The parameters $J_s$ and $J_k$ are the
exchange and
Kondo coupling constants, respectively.  In this study 
we ignore mesoscopic fluctuations,
taking equally spaced single-particle levels
[covering a band of width $2D=(N\!-\!1)\times\delta$]
and $\phi_\mu(0)=1/\sqrt{N}$. We also assume half-filling
of the band so that the number of dot electrons is $N$.
The average local density of states of the dot is $\rho=1/(N\delta)$
\cite{kaul05}.

The spin-rotation invariance of the Hamiltonian (\ref{eq:1}) implies
the conservation of the total spin
${\bf S}_{\rm tot}={\bf S}_d+{\bf S}_K$, so that $S_{\rm tot}$ and
$S_{\rm tot}^z=M_{\rm tot}$ are good quantum numbers.
To take advantage of this symmetry, we construct a good total spin basis by
coupling the eigenstates of the UH with
those of the Kondo spin. The UH eigenstates with dot spin $S_d$ are
characterized by $|\gamma S_d M_d \rangle$
($\gamma$ denotes orbital occupations $n_\mu$ and other quantum numbers
distinguishing between states of the same dot spin $S_d$).
Thus a basis of the coupled system with good total spin is $|\gamma S_d
S_{\rm tot} M_{\rm tot} \rangle$ (for simplicity
the quantum number
$S_K\!=\!1/2$ is omitted). In this basis the UH is diagonal with
energies $\sum_\mu \epsilon_\mu n_\mu\!-\!J_s S_d(S_d+1)$. The Kondo
term $H_{K}=J_k\,{\bf
     S}_{K}\cdot{\bf s}_{d}(0)$ is a scalar product of vector operators in
the uncoupled spaces. Thus, its matrix elements in the coupled basis conserve
$S_{\rm tot},M_{\rm tot}$ and are given by
\begin{multline}
\langle\gamma'S'_d S_{\rm tot} M_{\rm tot}|H_{K}|
\gamma S_d S_{\rm tot} M_{\rm tot}\rangle\!=\! J_k
(-1)^{S_{d}+1/2+S_{\rm tot}}\\\times\sqrt{\frac{3}{2}}\left\{\begin{array}{ccc}
S'_d & 1 & S_d \\ 1/2 & S_{tot} & 1/2  \end{array}\right\}
(\gamma'S'_d\parallel {\bf s}_d(0) \parallel
\gamma S_d) \,,\hspace{0.0cm}\label{eq:3}
\end{multline}
in terms of a Wigner-6j symbol and the reduced matrix element of the
spin density ${\bf s}_d(0)$ (known in closed form \cite{tureci06}).
In this formulation, the full Hamiltonian $H$ has a block
diagonal structure in $S_{\rm tot}, M_{\rm tot}$.

The problem is further simplified by transforming to the basis of sites $i$
($0\!\le\!i\!\le\!N\!-\!1$), in which the one-body part of the dot's
Hamiltonian is tridiagonal and  $H_K=J_k {\bf S}_K \cdot {\bf s}_0$
with ${\bf s}_i$  the
spin at site $i$ \cite{hewson93} (see Fig.~\ref{fig:1}b). The exchange
interaction is invariant under such transformation and has the same form as in
Eq.~(\ref{eq:1}) with ${\bf S}_d =\sum_{i=0}^{N-1} {\bf s}_i$. We can thus
recast our formalism in this site basis,
where only neighboring sites are coupled and the
Kondo spin interacts solely with site $0$. Due to these features
the many-body Hamiltonian matrix
in the site basis is more sparse than in the orbital
basis, allowing for an efficient diagonalization
in each subspace of good $S_{\rm tot}$
using a Lanczos-Arnoldi algorithm.
In this approach we can
conveniently diagonalize (\ref{eq:1}) for dots with up to $N \sim 12$
levels, where the total Hilbert space contains  $\sim 5.4\times 10^6$
basis states. 

\begin{figure}[!t]
\includegraphics*[angle=0,width=85mm]{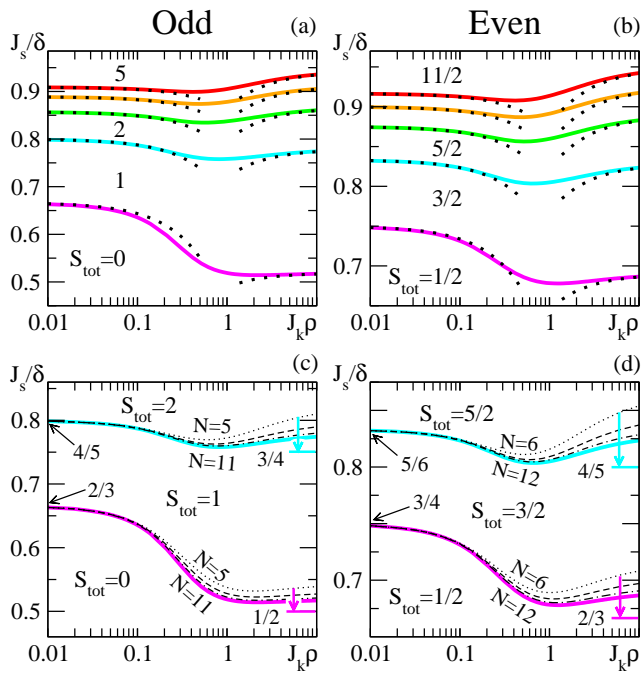}
\caption{(Color online) Ground-state spin $S_{\rm tot}$ of the
system in Fig.~\ref{fig:1} at finite exchange $J_s$ and Kondo coupling
$J_k$ for an odd (left column) and even (right column) number of
electrons $N$.\\(a),(b) Transition curves for $N\!=\!11$ (left) and
$N\!=\!12$ (right), separating regions of fixed $S_{\rm tot}$.
Numerical results (colored solid lines)
are compared with analytical estimates in the weak and strong
coupling limits (dotted lines).
(c),(d) Spin transition curves for fixed band width $2D$ but
different $N$ (top to bottom): $N\!=\!5,7,9,11$ (left) and
$N\!=\!6,8,10,12$ (right). For
increasing $N$ (at $J_k\rho\!\gg\!1$), the curves
converge to the Stoner staircase of a dot with $N\!-\!1$
electrons (colored arrows).}
\label{fig:2}
\end{figure}

We calculated the lowest many-body energy eigenvalue for each value of $S_{\rm
    tot}$ and thereby determined the ground-state value of the total spin
for different values of $J_s$ and $J_k$.
This quantity has been studied theoretically \cite{murthy05,kaul06}
and can be probed experimentally
\cite{parallel}. As $J_s$ increases, the ground state spin $S_{\rm
tot}$  is expected to undergo successive transitions to higher values
(known as the \emph{Stoner staircase}) until the dot becomes fully
polarized at $J_s\sim\delta$. For  $J_k\rightarrow 0$, the spin
transitions occur at $J^m_s=\delta(m+1)/(m+2)$, with $m=1,3,5,\dots$
($m=2,4,6,\dots$) for an odd (even) number of dot
electrons $N$. These transition steps in the Stoner staircase are shifted by
    the Kondo interaction. In Fig.~\ref{fig:2}a,b we show the transition curves
    (colored lines) separating regions of fixed ground-state spin $S_{\rm tot}$
    in the two-dimensional parameter space of $J_s,\,J_k$. We observe
that these curves
    are monotonically decreasing for $J_k\rho\lesssim 1$ and monotonically
    increasing for $J_k\rho\gtrsim 1$. In the strong-coupling limit, they
    converge to values of $J_s$ that are either lower (for
smaller $S_{\rm tot}$) or higher (for larger $S_{\rm tot}$) than their
corresponding weak-coupling values $J_s^m$.

To gain insight into the behavior of the transition curves, we evaluate them
for weak Kondo coupling
in first-order (degenerate) perturbation theory.
We find
\begin{equation}
J_s^m=[(m+1)\delta-\alpha_m J_k]/[m+2]\,,\label{eq:4}
\end{equation}
where $\alpha_m$ are positive constants of order one.
These perturbative results (dotted lines in Fig.~\ref{fig:2}a,b) agree well
with the numerical calculations for $J_k\rho \lesssim 0.1$.

The negative slope of the transition curves at weak coupling can be
understood by considering that in this weak-coupling regime
the Kondo spin ${\bf S}_K$ plays the role of an effective magnetic
field, polarizing the dot in the direction opposite to its own spin. This will
favor larger values of $S_d$ and thus also larger values of $S_{\rm
tot}$, hence the negative slope.  As we approach the Stoner
instability $J_s \to \delta$, (i.e., increasing $m$), the gain in
exchange correlations $ J_s S_d(S_d+1)$ dominates over the gain in
Kondo correlations, hence the flattening of the slope in this limit.

A perturbative analysis can also be carried out for the limit of
strong Kondo coupling, which, at zero temperature, is characterized
by $T_K\gg\delta$. For dots with sufficiently large $N$, this limit
can be reached already for $J_k\rho\ll 1$, where the Kondo
temperature is of the order $T_K\sim D\,e^{-1/J_k\rho}$ (at $J_s=0$)
\cite{hewson93}. However, for the present case of $N \lesssim 12$,
the limit $T_K\gg\delta$ requires $J_k\rho\gtrsim 1$ or,
equivalently, $J_k\gtrsim 2D$. The latter condition represents the
\emph{bare} strong-coupling
limit for which a perturbative solution is available (without
renormalizing the band width) \cite{nozi}. In this limit the Kondo
spin ${\bf S}_K$ and the spin  ${\bf s}_0$ at site 0 form a strongly
bound singlet $S_{K,0}=0$ (${\bf S}_{K,0}={\bf S}_K +{\bf s}_0$) that
is effectively decoupled from the rest of the spin-chain with sites
$i\ge 1$. The tridiagonal one-body Hamiltonian of sites
$1\le i\le N-1$ can be rediagonalized to give $\sum_{\mu\sigma}
\bar{\varepsilon}^{{\phantom \dagger}}_\mu \bar c^\dagger_{\mu\sigma}
\bar c^{{\phantom \dagger}}_{\mu\sigma}$, describing a ``reduced'' dot with
new orbital wave functions $\bar\phi_\mu({\bf r})$ and
single-particle energies $\bar{\varepsilon}_\mu$. This dot has one
less level and one less electron than the original dot. While the
original dot levels are equally spaced, the
level spacings in the reduced dot are given by
$\bar\varepsilon_{\mu+1}-\bar\varepsilon_\mu\equiv\delta_\mu\approx\delta +
\beta_\mu/N>\delta$, ($\beta_\mu>0$ are of order 1 and increase
monotonically from the new band center towards the band edges).

To explore how the strong-coupling limit is modified in the presence of
exchange
interaction in the dot, we rewrite the latter as $-J_s{\bf S}^2_d=-J_s{\bf
\bar S}_d^2- 2 J_s\, {\bf s}_0\cdot{\bf \bar S}_d -3 J_s/4$ where
${\bf \bar S}_d=\sum_{i=1}^{N-1} {\bf s}_i$
is the spin of the reduced dot. The cross term $-2 J_s\,{\bf
s}_0\cdot{\bf \bar S}_d$ has vanishing matrix elements in the singlet
subspace $S_{K,0}=0$ but induces virtual transitions to the triplet
subspace $S_{K,0}=1$ that renormalize the exchange coupling constant
$J_s\rightarrow\bar{J}_s = J_s(1+J_s/J_k)$ (details will be presented
elsewhere). To lowest order
in $1/J_k$, our system is thus described by an effective Hamiltonian
$\sum_{\mu\sigma}
\bar{\varepsilon}^{{\phantom \dagger}}_\mu \bar c^\dagger_{\mu\sigma}
\bar c^{{\phantom \dagger}}_{\mu\sigma} - \bar J_s{\bf \bar S}_d^2$
that has the form of a UH for the reduced dot with single-particle energies
$\bar{\varepsilon}_\mu$ and exchange constant $\bar J_s$
($\bar{J}_s\!\rightarrow \!J_s$ for $J_k\!\rightarrow\!\infty$). The spin
transition curves of this reduced dot (dotted lines in Fig.~\ref{fig:2}a,b)
are found to be in good agreement with the exact numerical curves when $J_k
\rho \gg 1$. Despite their accuracy in the appropriate limits, the
perturbative estimates do fail in the fully Kondo correlated
intermediate regime and our exact numerical solutions
are essential to show that there is a smooth crossover between the limits.

The spin transition curves in the crossover from weak to strong coupling
(see Fig.~\ref{fig:2}) are determined by two counteracting effects: (i) The
effective removal of an \emph{electron} from the dot shifts down
the Stoner staircase according to $J_s^m\rightarrow J_s^{m-1}$.
Since the reduced dot has one less electron, the shifted Stoner
staircase is associated with the opposite number parity of electrons.
(ii) The effective removal of a \emph{level} from the dot stretches
the step size in the staircase due to the larger level spacing in the
reduced dot (i.e.,
$\delta_\mu>\delta$), and thus increases the spin-transition values of
$J_s$. The downward shift in (i) is independent of $N$, but weakens for
increasing $m$ (where the step values $J_s^{m}$ are more densely spaced).
The upward shift in (ii) is a finite-size correction  $\sim 1/N$ that
decreases with $N$, but increases with $m$ because of the non-uniform
$\delta_\mu$. For smaller values of $J_s$, effect (i) dominates over
(ii), resulting in an overall downward shift of the transition values
in the strong-coupling limit (as compared to the weak-coupling values
$J_s^m$).
Close to the Stoner instability, however,  finite-size effects (ii)
dominate over (i), leading to transition values larger than $J_s^m$.

To investigate the interplay between effects (i) and (ii) more
closely, we compare the spin transition curves of our original
systems ($N=11,12$) with systems of equal band width $2D$, but
different values of $N$. Results shown in Fig.~\ref{fig:2}c,d
demonstrate that finite-size effects (ii) decrease with increasing
$N$, leading to a convergence of the strong coupling transition
curves towards $J_s^{m-1}$.

Dots with a large band width ($D\gg\delta$) could, in principle,
be studied by extending the NRG method to
include an exchange interaction. Truncation of the band width below $T_K$
leads to a strong-coupling effective Hamiltonian that includes additional
interaction terms. It would be interesting to investigate the effects of these
terms.

\begin{figure}[!t]
\includegraphics*[angle=0,width=80mm]{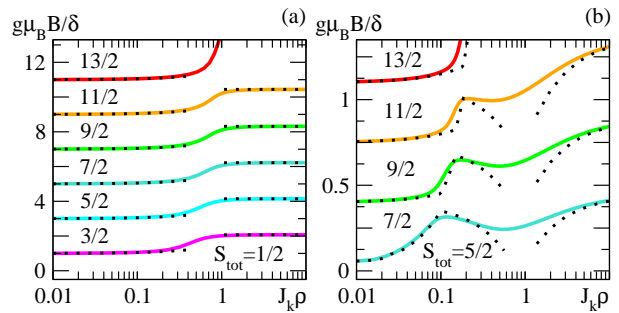}
\caption{(Color online) Spin transition curves for $N=12$ at Zeeman
     field $B$ and Kondo coupling $J_k$. The exchange coupling
     is (a) $J_s=0$ and (b) $J_s/\delta=0.825$. Colored solid (black dotted)
lines show the numerical (perturbative) solutions.}
\label{fig:3}
\end{figure}

Signatures of the interplay between the intra-dot exchange and the Kondo
coupling are revealed by applying an in-plane
field $B$ \cite{parallel}, adding a Zeeman term $g \mu_B
B\,S^z_{\rm tot}$
to the Hamiltonian $H$ in Eq.~(\ref{eq:1}) ($g$ is the gyromagnetic factor and
$\mu_B$ the Bohr magneton).  This term commutes with $H$
(although it breaks the $M_{\rm tot}$ degeneracy) and
favors a parallel configuration of dot spin ($\uparrow$) and
Kondo spin ($\Uparrow$), increasing $S_{\rm tot}$ at
$J_k=0$.  The addition of the Kondo interaction opposes
such a parallel alignment ($\uparrow\Uparrow$); correspondingly we 
find (Fig.~\ref{fig:3}a)
the spin transition values of $B$ monotonically increasing vs.~$J_k$.
A more complex and subtle behavior appears for $J_s \ne 0$  (see
Fig.~\ref{fig:3}b), where non-monotonic spin-transition curves
arise. The behavior in weak Kondo coupling can again be understood in 
perturbation
theory, for which the dot spin can still be regarded as a good quantum number.
In the limit $J_k\to 0$ the ground state will always be
the parallel configuration ($\uparrow\Uparrow$) aligned with the 
external field, so that
the spin transition lines slowly increase with $J_k$ (as for $J_s=0$).
At larger values of $J_k$ the energy of the anti-parallel
($\downarrow\Uparrow$) configuration
(with rearranged orbital occupancies) becomes lower in each spin 
subspace.  This happens
at lower $J_k$ for the lower value of
$S_{\rm tot}$, leading first to a marked increase in the slope of the 
transition line.
Increasing $J_k$ further one reaches the point at which the 
anti-parallel configuration is also favored in the subspace with 
higher $S_{\rm tot}$.  This decreases the 
slope of the spin transition line, making it negative in some cases. 
For even larger $J_k$ the perturbative picture 
breaks down and the transition curves
make a smooth crossover to the  strong
coupling picture where the effective exchange interaction constant
$\bar{J}_s$ now {\it decreases} with increasing $J_k$, favoring lower 
$S_{\rm tot}$ again
and  giving a positive slope to the transition 
curves.
Initial calculations with non-uniformly spaced single-particle levels
show that such a non-monotonic behavior persists in individual
dots. We defer detailed studies of mesoscopic fluctuations to future work.

The ground state spin of quantum dots (in the absence of Kondo
coupling) has been measured in a number of experiments
\cite{parallel}; typically this requires varying the
in-plane (Zeeman) field $B$, allowing us to map out
the spin transition diagram in the following manner. The
points of degeneracy between ground states with different spin for fixed $J_k$
are determinable by kinks in the slope of the Coulomb blockade peak
positions vs.~in-plane field $B$ \cite{baranger00}. Tuning $J_k$ (by means of a
pinch-off gate) at fixed $J_s$ will cause the kinks to
shift to higher or lower values of $B$ in a manner predictable from
our calculations.

In summary, the interplay of Kondo and ferromagnetic (Stoner)
interactions in large
quantum dots leads to interesting ground state
spin properties. The transitions to higher spin ground states due to
the Stoner interaction are shifted non-monotonically in the presence
of a Kondo spin.  At weak Kondo coupling the Kondo spin acts just as
an external field to assist ferromagnetic polarization. At strong
coupling the system is described (again) by a universal
Hamiltonian, but with a renormalized exchange constant for a reduced dot
with one less level and one less electron.  Ferromagnetic
polarization can be either enhanced or reduced in this limit,
depending on how close the dot is to the bulk Stoner instability.
The weak and strong coupling limits are described well by 
perturbation theory and our
exact numerical solutions find a smooth behavior in the
non-perturbative crossover region.

We thank S.~Adam, H.~Baranger,
L.~Glazman, M.~Kiselev, K.~Le Hur, C.~Marcus,
V.~Oganesyan, R.~Kaul, G.~Murthy, and S.~Schmidt for helpful discussions. This
work is supported by the Max-Kade Foundation, the W.M.~Keck Foundation,
U.S.~DOE grant No.~DE-FG-0291-ER-40608 and NSF grant DMR 0408636.


\begin{thebibliography}{02}

\bibitem{hewson93} A.C.~Hewson, {\it The Kondo Problem to Heavy Fermions}
(Cambridge Univ. Press, Cambridge, 1993).

\bibitem{leos} L.P.~Kouwenhoven and L.I.~Glazman, Phys.~World {\bf 14},
33 (2001).

\bibitem{goldhaber98} D. Goldhaber-Gordon {\em et al.},
    Nature {\bf 391}, 156 (1998); S.M.~Cronenwett,  T.H.~Oosterkamp,
L.P.~Kouwenhoven, Science {\bf 281}, 540 (1998).

\bibitem{glazman88} L.I.~Glazman and M.E.~Raikh, JETP Lett.~{\bf 47}, 452
(1988); T.K.~Ng and P.A.~Lee, Phys.~Rev.~Lett.~{\bf 61}, 1768 (1988);
Y.~Meir, N.S.~Wingreen, and P.A.~Lee, \emph{ibid.} {\bf 70}, 2601 (1993).

\bibitem{thimm99} W.B.~Thimm, J.~Kroha, and J.~von Delft,
Phys. Rev. Lett. {\bf 82}, 2143 (1999).

\bibitem{simon02} P.~Simon and I.~Affleck,
Phys.~Rev.~Lett.~{\bf 89}, 206602 (2002); P.S.~Cornaglia and
C.A.~Balseiro, \emph{ibid.} {\bf 90}, 216801 (2003).

\bibitem{kaul05} R.K.~Kaul {\em et al.},
Europhys.~Lett.~{\bf 71}, 973 (2005).

\bibitem{murthy05} G.~Murthy, Phys.~Rev.~Lett.~{\bf 94}, 126803 (2005).

\bibitem{kaul06} R.K.~Kaul {\em et al.},
Phys.~Rev.~Lett. {\bf 96}, 176802 (2006).

\bibitem{sindel07} J.~Martinek {\em et al.},
Phys.~Rev.~Lett.~{\bf 91}, 247202 (2003).

\bibitem{jalabert} R.A.~Jalabert, A.D.~Stone, and Y.~Alhassid,
Phys. Rev. Lett.~{\bf 68}, 3468 (1992); J.A.~Folk {\em et al.}, {\em ibid.}
{\bf 76}, 1699, (1996); A.M.~Chang {\em et al.}, {\em ibid.} {\bf 76}, 1695
(1996).

\bibitem{al00} Y.~Alhassid, Rev.~Mod.~Phys.~{\bf 72}, 895 (2000).

\bibitem{uniham} I.L.~Kurland, I.L.~Aleiner, and B.L.~Altshuler,
Phys. Rev.~B {\bf 62}, 14 886 (2000); I.L.~Aleiner, P.W.~Brouwer, and
L.I.~Glazman, Phys.~Rep.~{\bf 358}, 309 (2002).

\bibitem{rupp} Y.~Alhassid and T.~Rupp, Phys.~Rev.~Lett.~{\bf 91},
056801 (2003).

\bibitem{larkin72} L.~Shen, D.S.~Schreiber and A.J.~Arko,
Phys.~Rev.~{\bf 179}, 512 (1969);
M.J.~Zuckermann, Sol.~State Comm.~{\bf 9}, 1861 (1971);
A.I.~Larkin and V.I.~Melnikov, Zh. Eksp. Teor. Fiz.~{\bf 61} 1231 (1971)
[Sov.~Phys.~JETP {\bf 34}, 656 (1972)].

\bibitem{tureci06} H.E. T\"ureci and Y. Alhassid, Phys. Rev. B {\bf 74},
165333 (2006).

\bibitem{parallel}
D.S. Duncan \emph{et al.},
Appl.~Phys.~Lett.~{\bf 77}, 2183 (2000); J. A. Folk \emph{et al.},
Phys. Scr. {\bf T90}, 26 (2001);
L.P. Kouwenhoven,  D.G. Austing, and S. Tarucha, Rep. Prog. Phys. {\bf 64},
701 (2001);
S. Lindemann \emph{et al.},
Phys. Rev. B {\bf 66}, 195314 (2002); R.M. Potok \emph{et al.},
Phys. Rev. Lett. {\bf 91}, 016802 (2003);
R. Hanson \emph{et al.}, Rev. Mod. Phys. {\bf 79}, 1217
(2007).

\bibitem{nozi} P.~Nozi\`eres, J.~Low.~Temp.~Phys.~{\bf 17}, 31 (1974).

\bibitem{baranger00}
H.U. Baranger, D. Ullmo and L.I.
Glazman, Phys. Rev. B {\bf 61}, R2425 (2000).

%
\end{thebibliography}
\end{document}